# Rapid Stochastic Acceleration of Protons to Energies Above 100 TeV in the Accretion Column Of Hercules X-1

Paul A. Johnson*

Department of Physics, University of Leeds, Woodhouse Lane, Leeds, West Yorkshire, LS2 9JT, UK.

An investigation into the acceleration of protons by scattering off relativistic Alfvén waves in the accretion column of Hercules X-1 is presented. The mechanism is shown to achieve mean particle energies of 30 TeV under very reasonable assumptions about the environment, and 250 TeV is available under some circumstances. The highest individual energy attained is almost 1 PeV. The protons emerge in the form of a narrow beam directed at the inner edge of the accretion disk, which is favourable because of the reduced power requirement and presence of target material for $\gamma$-ray production.

astro-ph/9409012  6 Sep 1994

## 1. Introduction

In the last decade there have been several claims for high energy $\gamma-$ray emission between 250 GeV and 0.75 PeV from the X-ray binary pulsar Hercules X-1, some pulsed at the X-ray period [1] [2] [3], others at a period significantly blue-shifted by an amount agreed upon by several observers [4] [5] [6]. Recently, there have been several detections at a different blue-shifted period by another group [7]. There are reports of long term DC emission at a somewhat higher energy [8] [9], but these are not statistically very significant, and not corroborated by recent more sensitive searches [10][11]. The possibility that this and other such sources (e.g. Cyg X-3) are variable on timescales of several years is not surprising, given the evidence of the irregularity of the intrinsic power source in Her X-1, on a similar timescale, deduced from optical observations spanning the last century [12] [13]. However, here we will be concerned with the convincing pulsed behaviour at the slightly lower energy, which is characterised by episodic bursts of emission lasting typically 30 mins.

The progenitors of the $\gamma$-rays are usually assumed to be protons, because electrons would radiate their energy much too quickly in the magnetic field of the neutron star. In this case the $\gamma$-rays are thought to be produced by interaction of the protons in some gas target in the system, following the decay of any $\pi^o$ produced in collisions. If this picture is correct, then we run into problems with the huge implied proton luminosity, which appears to be higher than the total power available from accretion. [14]. For example, the flux from the Cygnus Array [6] implies a $\gamma$-ray luminosity of $1.6 \times 10^{38}$erg s$^{-1}$, if produced isotropically. The parent proton luminosity is probably an order of magnitude higher [15], implying a proton luminosity about 100 times higher than the estimated bolometric luminosity [16]. This energy budget problem is best balanced by invoking a narrow beam, which also has the advantage of explaining the episodic nature of the emission [14] if it wanders around in direction, possibly driven by the precession of the neutron star.

There are also problems in explaining the high energies that individual protons appear to be achieving. Observation of $\gamma$-rays up to energies of 750 TeV implies the acceleration of protons to an energy several times higher [15], probably several PeV. There have been several models proposed to explain the acceleration of protons to energies of a few PeV, but in general they have been shown to be untenable in the uniquely well understood environment of Her X-1. Shock acceleration above the pole of the neutron star [17] [18] fails because of the enormous magnetic field which inhibits the formation of scattering centres [19] [20]; the pul-sar wind shock mechanism [21] does not work because the pulsar spins too slowly [20]; the disk dynamo [22] [23] becomes very inefficient and the maximum energy available is much reduced for the high magnetic field found here [20]; turbulent reconnection of magnetic field may accelerate particles in X-ray binaries [24], but the total amount of energy available for particle acceleration is extremely low, and cannot explain the duration of the observed bursts [25]: the total amount of energy associated with the magnetic field ouside the Alfvén radius (where the plasma can still distort and wind-up the field) is much less than the energy requirement inferred from the high energy $\gamma$−ray observations.

---

*Now at the Department of Physics and Mathematical Physics, University of Adelaide, SA 5005, Australia

A pulsar dynamo model [26] successful in explaining $\gamma$-ray emission from rapidly spinning isolated pulsars has been extended to slower, accreting pulsars [27], but there may be problems in maintaining vacuum gaps in the presence of accreting plasma and the hot plasma of the accretion disk corona which is likely to form [28] [29].

The interaction between cosmic-ray protons and Alfvén waves has long been known to play an important role in cosmic-ray transport in the galaxy. Under certain circumstances, cosmic-rays generate Alfvén waves, and their velocity is limited to that of the waves. In other situations the cosmic-rays can be efficiently scattered by Alfvén waves, and in the presence of both forward and backward waves, Fermi acceleration can occur [30]. Such acceleration by fast-moving Alfvén waves in the magnetosphere of accreting neutron stars is a possible method of achieving proton energies of 1 PeV [31], and here we investigate the application of this mechanism to the accretion column of Her X-1. The preliminary results of this work have been described earlier in less detail [32]. The fluctuations which propagate as waves can be caused by the accretion splash on the neutron star, and travel outwards. Instabilities at the Alfvén radius can also create disturbances which can travel inwards. Reflection of protons by the converging field lines provides an opportunity for additional energy gain, as mirroring traps the particle in the system for longer, together with a focussing mechanism by which the protons emerge in a narrow range of angles.

The advantages of acceleration in the accretion column itself have been discussed elsewhere [14]. Since this work was first reported, Smith, Katz and Diamond [33] have extended their work to accretion columns in general, and although the approach is different from that given here, the results regarding maximum energy are in agreement.

## 2. The Environment

Her X-1 is an accreting X-ray binary pulsar whose environment is well understood. It lies 5.8 kpc distant [12], and consists of a neutron star spinning with a period $P_n = 1.24$ s, which orbits its companion every 1.7 d [34]. The neutron star is surrounded by a tilted and precessing accretion disk [35]. Her X-1 is unusual in that the mass of the primary is well measured, $M_n = 1.3 M_\odot$ [36], and is consistent with that expected for neutron stars [37] [38], for which a radius of $R_n = 10$ km is typical. Cyclotron lines in the X-ray spectrum indicate the presence of a magnetic field at the pole of the neutron star with flux density $B_o = 5 \times 10^{12}$ G [39], in line with expectation for such an object. Inside the Alfvén radius, $R_A$, which is coincident with the co-rotation radius at $\sim 200 R_n$, this strong magnetic field controls the flow of gas. The gas slides as an ionised spray [40], its geometry described by the magnetic field lines, and forms an accretion column, landing on the surface of the neutron star over an area of somewhat less than 1 km$^2$. In this investigation a simplified version of the accretion column is used as follows: Let $r$ be the distance from the centre of the neutron star, and $x$ the distance from the column centre. The column extends between $r = R_n$ to $r = R_A$ with circular cross sectional area $A = A_o(r/R_n)^3$, radius $x_c = (A/\pi)^{1/2}$, free fall gas velocity $v_g = v_o(r/R_n)^{-1/2}$ and magnetic field strength $B = B_o(r/R_n)^{-3}$. The zero subscript refers to the value at the neutron star pole: $B_o = 5 \times 10^{12}$ Gauss, and $v_o = \sqrt{2GM_n/R_n} = 1.3 \times 10^{10}$ cm s$^{-1}$ are known. The density at the pole is estimated from equipartition between the magnetic pressure and the ram pressure of the accreting gas at the Alfvén radius, and then applying continuity to give $\rho_o = 0.01$g cm$^{-3}$. The area of the column at the pole can be estimated from the disk thickness at the Alfvén radius to be $A_o = 0.03$ km$^2$ [25]. These figures yield a mass flux $\dot{M} = 4 \times 10^{16}$ g s$^{-1}$, and a total available gravitational power of $7 \times 10^{36}$ erg s$^{-1}$ *per pole*, which is consistent with the bolometric luminosity, $2 \times 10^{37}$ erg s$^{-1}$ [16]. The temperature of gas required to produce the observed luminosity from the polar area is $1.5 \times 10^8$K, in agreement with that measured [34].

Since the gas in the accretion column is expected to be fully ionised, Alfvén waves will propagate with a velocity $v_A$ given by

$$v_A = c/(1 + 4\pi\rho c^2/B^2)^{1/2} \qquad (1)$$

after [41]. The velocity of the Alfvén waves is usually much greater than the velocity of the infalling gas, and so the latter is treated as stationary in this investigation. The quantities $\rho$, $B$ and $v_A$ are assumed independent of the distance from the column centre, $x$.

### 2.1. Spectrum of Alfvén Waves

Two modes of Alfvén waves are considered: one mode moving outward in the $+r$ direction and one moving inward in the $-r$ direction; these will be referred to as the '+' and '−' modes, and both have the same $v_A = v_A(r)$. The amount of power passing through the column in the form of the two modes of Alfvén waves is denoted $L_+$ and $L_-$. Here we assume that the '+' mode originates at the neutron star surface, the waves in the field perhaps caused by the accretion splash. The smallest wavenumber for this mode is taken to be $k_{o+} = 10/R_n$. The '−' mode is assumed to start at the Alfvén radius, the waves possibly initiated by the Rayleigh-Taylor instabilities at the interface between orbital motion in the disk and free fall motion in the accretion column [42]. The smallest wavenumber for this mode is taken to be $k_{o-} = 10/R_A$. In other words, the longest wavelength in each spectrum is approximately one tenth the size of the region causing the disturbances in the field lines. These minumum wavenumbers are uncertain, and are varied to assess the effect on the acceleration process.

To calculate how strongly a particle of a given energy is scattered, we need to know how the energy density is distributed over the spectrum of Alfvén waves. We will assume that the spectrum takes the form

$$\frac{d(b^2)}{dk} = \xi k^n \qquad (2)$$

where $b$ is the amplitude of the waves at a given wavenumber, $k$. Hence we can find the total energy density resident in the spectrum, $U$, by integrating between $k_1$ and $k_2$ which, if $k_2 \gg k_1$ and $n < -1$ reduces to

$$U = \frac{-\xi k_1^{n+1}}{8\pi(n+1)}. \qquad (3)$$

Now if we specify that the total magnetic energy flux in Alfvén waves to be a constant $L$ at any distance from the neutron star, then since $L = U v_A A$ we have

$$\xi = \frac{-8\pi(n+1)L}{A v_A k_1^{n+1}} \qquad (4)$$

for any assumed $L$, $n$ and $k_1$. We also have to allow for the fact that the velocity of the wave, $v_A$ changes with $r$, and hence a wave with initial $k_1 = k_o$ and $v_A = v_{A,o}$ (at the source of the spectrum) will become $k_1 = k_o v_{A,o}/v_A$ as $v_A$ changes. Finally we have

$$\xi = \frac{-8\pi(n+1)L}{A v_A (v_{A,o} k_o/v_A)^{n+1}} \qquad (5)$$

which from equation 2 gives us $d(b^2)/dk$ for each mode as a function of $v_A$, once $L$, $n$ and $k_o$ are specified, $A$ having being found in section 2. Of the variables, the total power in the Alfvén waves, $L$, is the hardest to determine.

### 2.2. Electric and Magnetic Fields Associated with Alfvén waves

Consider an Alfvén wave propagating in a plasma along the direction of a strong ambient field. Seen in the frame in which the plasma has no bulk motion, the total magnetic field is $(\mathbf{B} + \mathbf{b})$, where $\mathbf{b}$ is an oscillating component orthogonal to the ambient field $\mathbf{B}$. If the conductivity is very high, there is also an oscillating electric field $\mathbf{E}$ which is in phase with $\mathbf{b}$, and orthogonal to $\mathbf{B}$ and $\mathbf{b}$, travelling with a velocity $v_A$ such that $E = v_A b$. The velocity of the wave relative to the plasma is given by equation 1 (At low speeds, when $\rho/B^2$ is high enough, this expression reduces to the familiar $B/(4\pi\rho)^{1/2}$). The electric and magnetic field may be found for any other frame moving at a velocity $v$ along **B** by using the Lorentz transforms. In the rest frame of the wave, moving at a velocity $v_A$ along **B**, the transverse fields are

$$E' = \gamma_A(E - v_A b) = 0 \qquad (6)$$

and

$$b' = \gamma_A(b - v_A E/c^2), \qquad (7)$$

where is $\gamma_A$ is the Lorentz factor of the waves. Similarly the inverse transforms become

$$E = \gamma_A v_A b' \qquad (8)$$

$$b = \gamma_A b'. \qquad (9)$$

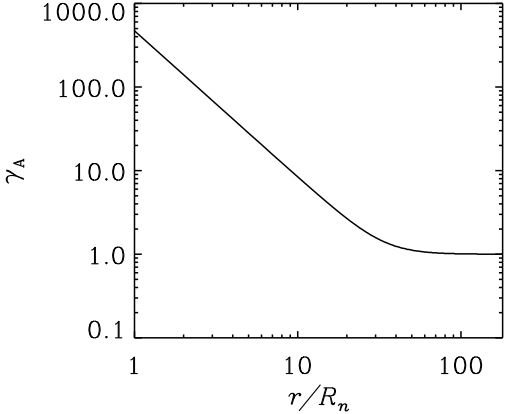

Figure 1. The Lorentz factor of the Alfvén waves, $\gamma_A$, as a function of the distance from the neutron star, $r$, in the accretion column. The distance $r$ is given in units of the neutron star radius, $R_n$.

The behaviour of $\gamma_A$ in the accretion column is shown in Fig. 1. It can be seen that in the inner half of the accretion column, the waves become relativistic. In the extreme limit, the electric field associated with the wave is $E = cb$, and the Alfvén wave takes on the form of an electromagnetic wave in free space. The circumstances under which this transition has occurred are unusual: the condition $B^2/\rho \gg 4\pi c^2$, obtained from Eqn. 1, is normally satisfied because $\rho \to 0$, but here the density is actually increasing. It is the rapid rise of $B$ which causes the transition in this case.

## 3. Calculation of Scattering

The scattering of protons by Alfvén waves depends on the total force on the particle. For slowly moving Alfvén waves, the magnetic field only need be considered. For relativistic waves, the electric field becomes appreciable, and the magnetic and electric forces become comparable. In this investigation the scattering is calculated for the particle in the frame of the wave. Here, the electric field $E'$ is zero, and only the magnetic field $b'$ remains. Once the magnetic field in the frame of the column, $b$, is known, that in the frame of the wave, $b'$, is found from equation 9. The calculation of the scattering in this frame then proceeds as described in the next section.

### 3.1. Pitch Angle Scattering Due to Alfvén Waves

Consider a particle spiralling in a stationary magnetic field of strength $B$, so that the particle has a Larmor radius $r_L = p/eB$, pitch angle $\theta$, and hence pitch equal to $2\pi r_L \cos\theta$. Suppose a spectrum of small scale tranverse waves is superposed. At a particular time, at the location of the particle, this transverse field will be of magnitude $b$. In a time $\Delta t$ the particle moves a distance $s = c\Delta t$, and the pitch angle changes by

$$\Delta\theta = \frac{s}{r_L}\frac{b}{B}. \qquad (10)$$

The total effect on the particle depends strongly on wavelength, $\lambda$: for relatively short wavelengths, such that $\lambda \ll r_L$, the waves add incoherently to give a very erratic transverse field. We can assume deflections are uncorrelated over even quite short distances, and the effective field, $b_{eff} \propto \sqrt{s}$. At the other extreme of wavelength, when $\lambda \gg r_L$, $b$ does not change appreciably as the particle turns round its helix. Effectively the total field is changed gradually, without an appreciable change in magnitude, and the guiding centre will follow the disturbed field line adiabatically. Here we have no change in pitch angle, very little displacement (unless very long wavlength waves cause field line wandering) and hence no cross-field diffusion.

However, when $\lambda \simeq r_L$, the particle sees a constant transverse field all the way round its orbit, which causes a large change in pitch angle. If we decompose the wave into right and left circularly polarised components, one set will have a radial component of $b$ which turns with the particle and so gives a constant $d\theta/dt$, and hence $\Delta\theta \propto s$ instead of $\sqrt{s}$. In fact, integrating over the whole spectrum

$$\Delta\theta_{rms} = \frac{b_{eff}}{B}\sqrt{\frac{s}{r_L}} \qquad (11)$$

such that $\Delta\theta \propto \sqrt{s}$ for scattering by waves where $\lambda < r_L$ and zero for longer wavelengths. The effective field, $b_{eff}$ is given by

$$b_{eff} = \sqrt{\pi K \left(\frac{d(b^2)}{dk}\right)_{k=K}}, \qquad (12)$$

where $K = 1/2\pi r_L \cos\theta$ is the inverse of the pitch, or the effective wavenumber of the particle. These last two equations are equivalent to the treatment of [30] which was more hydrodynamical in nature.

The cross-field displacement is also important when $\lambda < r_L$, and in fact the guiding centre is displaced in a random azimuthal direction by an amount

$$\Delta x_{rms} = r_L \Delta\theta \sin\theta. \qquad (13)$$

This is necessary to determine whether a particle wanders sideways out of the accretion column.

### 3.2. Pitch Angle Change Due to a Converging Field

The pitch angle is also changed, this time systematically, by the converging magnetic field of the neutron star. If the proton is near the magnetic axis at a distance $r$ from the neutron star, then $B \propto r^{-3}$, and hence

$$\frac{dB}{B} = -3\frac{dr}{r}. \qquad (14)$$

From the adiabatic invariant $\sin^2\theta \propto B$, if the proton travels a distance $s$ we have by differentation with respect to $s$

$$\frac{1}{B}\frac{d(\sin^2\theta)}{ds} - \frac{\sin^2\theta}{B^2}\frac{dB}{ds} = 0 \qquad (15)$$

whence from $dr/ds = -\cos\theta$ and equation 14,

$$\frac{d\theta}{ds} = \frac{3}{2}\frac{\sin\theta}{r}. \qquad (16)$$

Of course, no displacement of the guiding centre occurs in this case.

### 3.3. Transformations and particle energy gain

The mechanism for energy gain is similar to second-order Fermi acceleration, except that here the scattering centres are relativistic Alfvén waves, which provides an opportunity for very rapid acceleration. An Alfvén wave of amplitude $b$ travelling with velocity $v_A$ and Lorentz factor $\gamma_A$ in the accretion column frame has amplitude $b' = b/\gamma_A$ in the frame of the Alfvén waves. In transforming to the wave frame a particle with energy $\varepsilon_i$, velocity $\beta c$ and initial longitudinal momentum $p_i$ in the column frame becomes

$$p'_i = \gamma_A(p_i - \beta_A \varepsilon_i) \qquad (17)$$
$$\varepsilon'_i = \gamma_A(\varepsilon_i - \beta_A p_i) \qquad (18)$$

in the frame of the wave (quantities primed). After accounting for the scatter of the particle we use

$$p_f = \gamma_A(p'_f + \beta_A \varepsilon'_i) \qquad (19)$$
$$\varepsilon_f = \gamma_A(\varepsilon'_i + \beta_A p'_f) \qquad (20)$$

to transform back to the column frame, using the subscripts $i$ and $f$ to denote the initial and final states. The energy remains unchanged in the frame of the wave, $\varepsilon'_i = \varepsilon'_f$, and the change in $p'$ depends on the change in pitch angle. In fact for total reflection, where $p'_f = -p'_i$, $\Delta\varepsilon/\varepsilon = 2\gamma_A^2\beta_A(\beta + \beta_A)$, which reduces to $\Delta\varepsilon/\varepsilon = 4\gamma_A^2$ for relativistic protons and waves. From the high values of $\gamma_A$ attained in the accretion column as indicated in Fig. 1, it is evident that we might expect large energy gains: $\gamma_A \sim 100$ implies $\Delta\varepsilon/\varepsilon \sim 4 \times 10^4$. Such energy gains can only be achieved if the irregularities in the field are strong enough to be able to turn the particle round in the space available, and if the particle is not to wander sideways outside of the column.

## 4. The Simulation

Protons are injected isotropically into the accretion column of section 2 with initial energy $\varepsilon_i$ and position $r_i$. A transformation is then made into the frame of one of the two Alfvén wave modes. Here the particle is moved by a small amount $s$, and the pitch angle scattering and sideways displacement are calculated. The amount of pitch angle scattering is determined by the method of rejection, by sampling a gaussian distribution with width given by $\theta_{rms}$ in equation 11. This displacement is optimised to correspond to a reasonable pitch angle change so that the particle is not followed to undue precision. The contribution to the scattering from the other wave mode is then calculated, and then finally, back in the frame of the plasma, the adiabatic pitch angle change is accounted for. This process is repeated whilst keeping track of the particle position and momentum, until the particle either leaks from the side of the column ($x > x_c$), hits the star ($r < R_n$) or escapes ($r > R_A$). Many protons are thus injected to build up a spectrum of final energies for each of the three possible fates.

## 5. Results

### 5.1. The Canonical Spectrum

The canonical parameters for this investigation are those mentioned in section 2 together with $n = -3/2$, $L_+ = L_- = 10^{35}$ ergs$^{-1}$, $r_i = R_A/2$ and $\varepsilon_i = 1.1$ GeV. The power in the two modes of Alfvén waves is set at roughly 1% of the total gravitational power available per pole ($L_b/2$), which at the Alfvén radius, $R_A$, corresponds to an energy density $U \simeq U_B$, the background magnetic energy density, and hence represents a realistic upper limit to the power available. At the pole of course, the energy density of the waves is far less than that in the ambient field, $U \simeq 10^{-8} U_B$. Concerning the other parameters, the spectral index of the magnetic irregularities corresponds to a Kolmogorov spectrum and the initial energy of the particles is approximately equal to the energy that the protons achieve in free-fall.

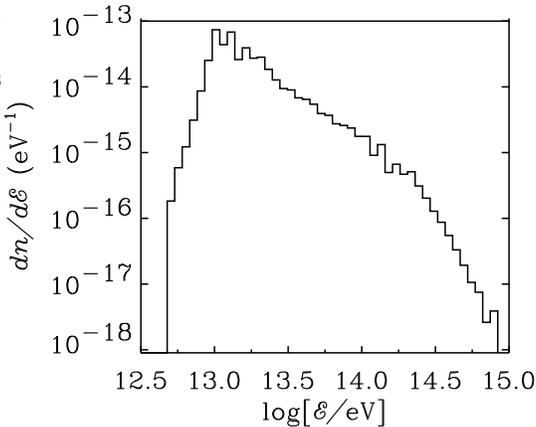

Figure 2. The spectrum of protons which emerge at the Alfvén radius after acceleration in the accretion column.

The proton spectrum per injected proton, $dn/d\varepsilon$, obtained for these parameters is shown in Fig. 2. The majority of protons in fact emerge from the column at $R_A$, only a small fraction leaking sideways and very few hitting the star. Most protons emerge with more than 10 TeV of energy, and the mean energy obtained is 30 TeV. The spectrum extends up to around 250 TeV with a slope of -2.0, after which point it steepens to around -3.0. Above this steepening there is no evidence for an absolute cut off. The statistics limit the maximum proton energy attained in this investigation to just under 1 PeV.

The energy and position of a typical proton are shown as a function of time in Fig. 3. This particular proton achieves an energy of 30 TeV in 10 ms.

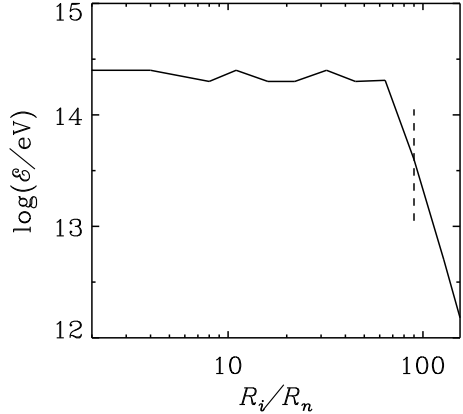

Figure 4. The mean escape energy, $\varepsilon$, versus the particle injection distance from the neutron star. The distance is given in units of the neutron star radius, $R_n$. The dashed line indicates the canonical value used in the investigation.

### 5.2. Sensitivity of spectrum to input parameters

The parameters $B_o$, $A_o$, $\rho_o$ and $v_o$ are assumed to be sufficiently well known that they are not altered in this investigation. The remainder are not well known, however, and each is varied in turn whilst holding the others at the canonical values. The emergent spectrum is obtained for a range of $r_i$, $n$, $k_{o+}$ and $k_{o-}$, $L_+$ and $L_-$.

The injection point, $r_i$, was varied between 2 $R_n$, near to the neutron star, out to near the Alfvén radius, $R_A$, and the corresponding mean escape energy is given in Fig. 4. The mean escape energy shows some dependence on $r_i$: the energy is degraded markedly as $r_i$ approaches $R_A$, but this is to be expected as the protons tend to escape before they have been significantly accelerated. The mean particle energy is enhanced to 200 TeV as the injection point moves a closer to the star than the canonical value, $R_A/2$. Therefore, if the protons had been injected uniformly along the column, the mean energy would have been nearer this value.

A range of $n$ from -1.0625 to -2 produced no dramatic changes in the spectrum. The spectrum is slightly degraded for the most negative values of $n$. More important seems to be the power contained in the Alfvén waves, $L_+$ and $L_-$, rather than the shape of the spectrum. The range of power tried for these parameters spans $10^{25}$ to $10^{30}$ W in view of the uncertainty in the mechanism which mechanically couples the inflowing gas to the field lines to produce Alfvén waves. The mean energy obtained as a function of the power residing in the two Alfvén waves modes is illustrated in Fig. 6. The spectrum is not enhanced significantly when both $L_+$ and $L_-$ are above the canonical values. However, as $L_-$ drops there is a degradation in energy, and a tendency for more protons to leak from the column before emerging at $R_A$. If $L_+$ is lowered, the mean energy of the protons drops somewhat faster. However, the energy degradation is severe only for very small values of $L$: remarkably, even with just $10^{26}$ W in each of the modes ($10^{-4}L_b/2$), the protons still emerge with a mean energy of nearly 3 TeV.

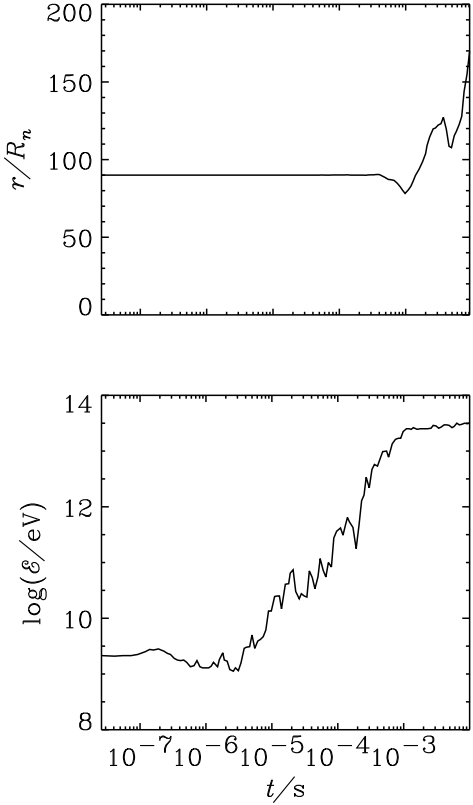

Figure 3. Some features of the history of a typical particle are shown. Both (a) the energy, $\varepsilon$, and (b) the distance from the neutron star $r$ are given as a function of time since injection, $t$. This particle escapes after 10 ms with an energy of 30 TeV.

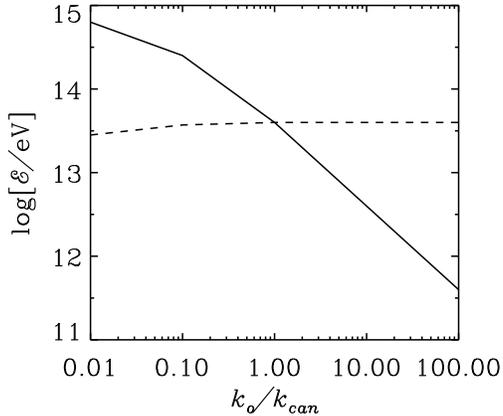

Figure 5. The mean escape energy, $\varepsilon$, versus the minimum wavenumber at the source of the Alfvén waves, $k_{o+}$ (full) and $k_{o-}$ (dotted), both given is a fraction of the canonical values described in the text.

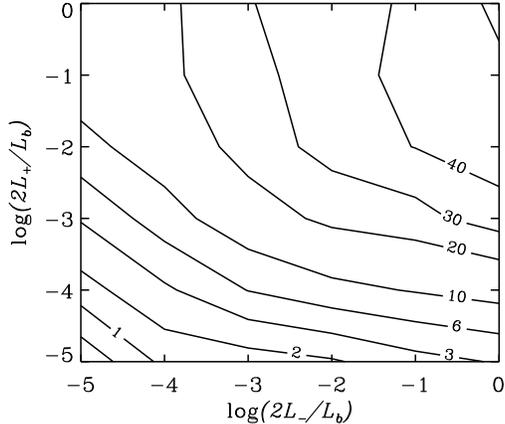

Figure 6. A contour map of the mean escape energy, $\varepsilon$, as a function of the power in both the in (-) and out (+) wave modes. The power is expressed as a fraction of the accretion luminosity per pole ($L_b/2$). The contour lines are labelled in units of TeV.

The cut-off wavelength present in the spectrum of Alfvén waves is hard to estimate, and so $k_{o+}$ and $k_{o-}$ are tried at factors of up to 100 either side of the canonical values, and the results given in Fig. 5. The spectrum is insensitive to $k_{o-}$, but the maximum achieved energy goes up dramatically as $k_{o+}$ is lowered. This effect might be expected: the maximum attainable energy for a particle is achieved when the gyroradius of the particle is comparable to the longest wavelength present in the spectrum of the Alfvén wave mode by which it is being scattered. For the '+' mode, the canonical value $k_{o+}$ is equivalent to a wavelength of $0.1 R_n$, which is the same as the gyroradius of a particle of energy 40 TeV at $R_A/2$. This is just below the break in the spectrum of accelerated protons. Particles which undergo scattering a little further in towards the star can reach the higher energies seen in the simulation. If $k_{o+}$ is decreased then this maximum energy is seen to rise as expected. For maximum wavelengths of $1\ R_n$ in the '+' mode, the mean energy is enhanced to above 100 TeV. In contrast, the wavelength associated with the incoming waves is al-ways large enough to interact with protons in the energy range of interest, and the mean energy of accelerated particles shows almost no change when this parameter is varied by a factor of a 100 either side of the canonical value.

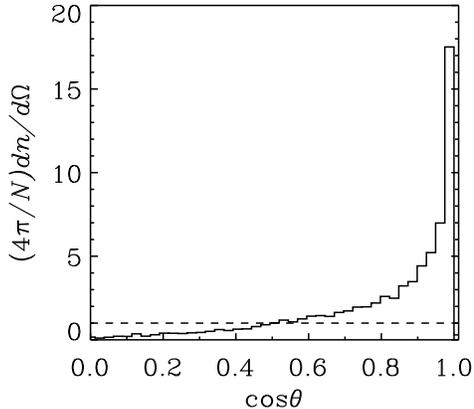

Figure 7. The distribution over solid angle, $dn/d\Omega$ for those protons which emerge at the Alfvén radius. The data is expressed as a fraction of that expected for an isotropic distribution (dashed line).

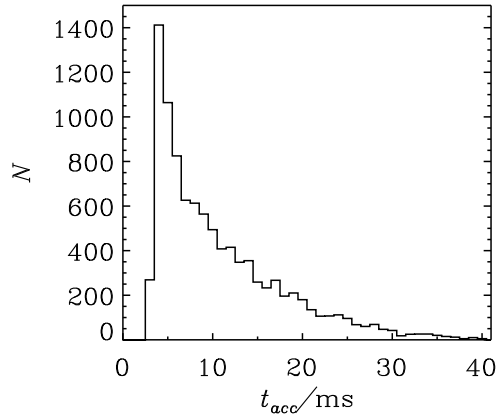

Figure 8. The distribution of acceleration times, $t_{acc}$, for those protons which emerge at the Alfvén radius.

The funnel-shape of the magnetic field in the accretion column has the important effect of systematically reducing the pitch angle of the particle through adiabatic invariance, which keeps it in the system for longer by inhibiting leakage out of the side of the column. More importantly, by this same process the protons are also focussed into a beam: by the time they emerge at $R_A$, the protons are travelling roughly parallel to the magnetic field, as shown in Fig. 7, where the distribution in solid angle is given, $dN/d\Omega$ together with that expected for an isotropic distribution. For those particles travelling with zero pitch angle along the accretion column the intensity is over 17 times higher than if the protons had emerged isotropically. The particle beam would stand the highest chance of being detected when directed at Earth, presumably following the production of $\gamma$-rays after passing through gas in the accretion disk. In this case, detection would therefore cause an overestimation of the power of the accelerator by this substantial factor, if the usual assumptions of isotropic emission are made.

For the canonical spectrum, the time spent in the system for each proton is shown in Fig. 8. Protons leave the system in their accelerated state after typically 10 ms. The process is therefore very fast compared to the timescale of the neutron star rotation period 1.24s, and significantly faster then the infall time of the accreting plasma from the Alfvèn radius to the neutron star,

$$t_{infall} = \left(\frac{2R_A^3}{9GM_n}\right)^{1/2} \text{s}, \qquad (21)$$

which is approximately 90 ms.

## 6. Loss Mechanisms

In the simulation, no energy losses were assumed. Here we check various possible methods of energy loss which could restrict the maximum energy attained. The protons are subject to energy losses by synchrotron radiation in the strong magnetic field of the neutron star, interaction with X-rays from the hot-spot, and inelastic collisions with the gas in the accretion column.

## 6.1. Losses through $pp$ collisions in accretion column

The density of gas in the accretion column, $\rho$ is given in section 2. A proton travelling from a distance $r$ from the neutron star, outwards through the accretion column will traverse an amount of matter $\Sigma$ given by

$$\Sigma = 6600 \left(\frac{r}{R_n}\right)^{-3/2} \text{ g cm}^{-2} \qquad (22)$$

Protons travelling from a point $r/R_n = 20$ axially outwards will thus traverse 75 gcm$^{-2}$ of matter. However, the protons are accelerated in a region further out than this, and they may encounter substantially less than this amount. On the other hand, the protons do not travel directly out of the system. In fact, from the simulation, the distribution of the amount of matter traversed is shown in Fig. 9. Typically less than 15 g cm$^{-2}$ is encountered (which indicates that the protons get no closer than about 60 $R_n$ to the neutron star). This is considerably less than the $pp$ interaction mean free path, $\sim 80$ g cm$^{-2}$ and so the accelerator will not suffer serious losses through this channel. However, this amount of matter may serve as a thin target for continuous weak $\gamma$-ray production.

## 6.2. Synchrotron Loss

The synchrotron loss time is given by

$$t_s = \frac{3m_p^3 c^5}{2e^4 B^2 \gamma} = 3 \times 10^{-6} B_{12}^{-2} E_9^{-1} \text{ s} \qquad (23)$$

where the proton is travelling orthogonal to the magnetic field [43]. If we define the mean free path for synchrotron loss as $\lambda_s = ct_s$, expressing it as a fraction of the distance from the neutron star, we get

$$\frac{\lambda_s}{r} = 3.6 \times 10^{-3} \left(\frac{r}{R_n}\right)^5 E_9^{-1}. \qquad (24)$$

assuming a dipole field, $B \propto r^{-3}$, and an average pitch angle of $\pi/4$. An indication of serious energy loss is given when $\lambda_s/r$ is less than unity i.e.

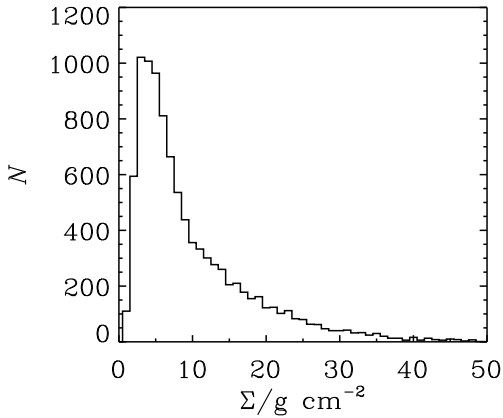

Figure 9. The distribution of the total grammage, $\Sigma$, encountered by those protons which emerge at the Alfvén radius.

the mean free path is less than the scale size of the region. This occurs for $r < 12R_n$ for 1 TeV protons, and $r < 50R_n$ for 1 PeV protons. Since the acceleration generally occurs in a region further out than this, and most of the time the protons are travelling with very small pitch angles, energy loss by this process is not important except possibly for the very highest energy protons when they get relatively close to the neutron star, in which case we might expect secondary $\gamma$-rays to be produced. In this way synchrotron loss could ultimately limit the maximum energy achievable by this acceleration mechanism to around 1 PeV.

## 6.3. Losses through photopion production

Photopion production has been shown [44] to be a possible source of $\gamma$-rays for protons above 1 PeV in X-ray binaries. The reaction $p + \gamma \rightarrow p + \pi_o$ has a threshold given by

$$\varepsilon_p \varepsilon_X > 1.5 \times 10^{17} (1 - \cos\alpha)^{-1} \text{ eV}^2 \qquad (25)$$

where $\varepsilon_p$ and $\varepsilon_X$ are the energy of the proton and photon respectively in eV, and $\alpha$ is the angle between them. Since the temperature of the hotspot on the neutron star is $10^8$ K, the average energy of the emitted photons is $<\varepsilon_X> \simeq 25$ keV, and so the threshold energy for this reaction is then 3 TeV for head on collisions, and rises to 400 TeV when $\alpha \simeq 10°$. As the protons are bouncing back and forth in the accretion column they will meet the X-ray flux at the full range of angles, and so this reaction is in the energy region of interest. Above this threshold the mean free path of a proton is given approximately by

$$\lambda_{p\gamma} = 600 \left(\frac{L_X}{10^{37}\text{erg s}^{-1}}\right) \left(\frac{r_x}{R_n}\right)^2 \varepsilon_X \text{ cm} \quad (26)$$

where $r_x$ is the distance from the hot-spot (rather than the distance from the centre of the neutron star). Substituting for $\varepsilon_X$ and the X-ray luminosity $L_X \simeq L_b = 2 \times 10^{37}$ erg s$^{-1}$, and expressing as a fraction of the radius at that point,

$$\frac{\lambda_{p\gamma}}{r} = 7.5 \left(\frac{r_x}{R_n}\right), \quad (27)$$

which is greater than unity except very close to the neutron star.

In addition, the protons will be screened from the X-ray flux by the gas in the column. The semi-thickness of the column, $\Sigma_w$ may be expressed as

$$\Sigma_w = \rho(r)x_c(r) = 100 \left(\frac{r}{R_n}\right)^{-1} \text{gcm}^{-2} \quad (28)$$

which is greater then the Thompson value of 2.5 gcm$^{-2}$ for the inner part of the column. Assuming firstly that the protons are concentrated in the centre of the accretion column, and secondly that the X-rays impinge at an angle normal to the column, the overall interaction length reduces to

$$\frac{\lambda_{p\gamma}}{r} = 7.5r \exp\left[40 \left(\frac{r}{R_n}\right)^{-1}\right], \quad (29)$$

which is significantly greater than unity for the whole length of the column. This is illustrated in Fig. 10, together with the interaction length in the case of no screening. The protons will probably be distributed more evenly in the column of course, but this will make little difference at large distances since even with no screening the interaction length there is still very large. Even if the column is curved overall, closer in it stands vertically above the hot spot, and the screening will be very much more effective because the X-rays will not be coming in the sides, but up through the lower, densest part of the column. For these reasons, the interaction length given in the Figure is probably conservative. Therefore, partly because of the screening effect of having the accelerator in the gas of the accretion column, significant loss by photopion production does not occur. As a consequence of this, secondary $\gamma$-ray production by this mechanism will also be suppressed. Photopair production will also occur with a similar energy loss rate, but at substantially lower energies than the pion production and so this channel is also unimportant.

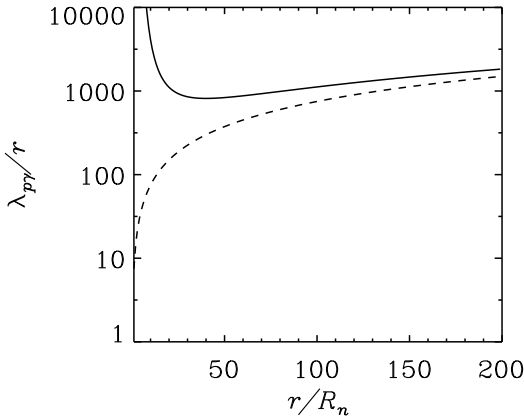

Figure 10. The $p\gamma$ interaction length $\lambda_{p\gamma}$ as a function of the distance from the neutron star, $r$. The interaction length is expressed as a fraction of $r$, and is shown with and without the screening effect of the gas in the accretion column (full and dashed lines respectively).

## 7. Estimate of $\gamma$-ray flux

The flux of $\gamma$-rays from Her X-1 detected by the various experiments can be summarised as follows. Around 1 TeV, long term fluxes of several $10^{-11}$cm$^{-2}$s$^{-1}$ and bursts up to $10^{-9}$cm$^{-2}$s$^{-1}$ are seen, whilst at several hundred TeV, long term fluxes around $10^{-13}$cm$^{-2}$s$^{-1}$ and bursts up to $10^{-11}$cm$^{-2}$s$^{-1}$ have been reported. We now consider the acceleration process described here in view of these fluxes.

Significant $\gamma$-ray production depends on the existence of a sufficiently dense gas target to act as a beam dump, and the exact flux depends on the details of the hadronic cascade which will be initiated when the protons strike such a target. Here we will simply assume that a fraction $\epsilon_\gamma \simeq 10\%$ of the proton luminosity is converted to $\gamma$-rays. The inner edge of the accretion disk may function as the target, and most of the time may be too dense or too thin to act so efficiently, but certainly some of the time this figure will serve as a good approximation.

After passing through such a target, the flux of $\gamma$-rays measured at Earth will be

$$F_\gamma = Q \frac{\epsilon_\gamma \epsilon_p \epsilon_A L_B}{4\pi D^2} \frac{1}{\bar{E}_p} f(> E_p) \quad \text{cm}^{2-}\text{s}^{-1} \quad (30)$$

where $Q$ is the enhancement due to proton beaming; $\epsilon_A$ and $\epsilon_p$ are the fraction of the bolometric luminosity converted to Alfven waves and then high energy protons repectively; $\bar{E}_p$ is the mean energy of high energy protons produced by the accelerator; $f(> E_p)$ is the fraction of protons emerging with energy greater than $E_p$; and finally, $D$ is the distance to the source. Using $Q = 17$, $\epsilon_\gamma = 10\%$, $D = 5.8$ kpc and $\bar{E}_p \simeq 30$ TeV we obtain

$$F_\gamma = 1.7 \times 10^{-10} \epsilon_p \epsilon_A f(> E_p) \quad \text{cm}^{2-}\text{s}^{-1}. \quad (31)$$

Assuming TeV $\gamma$-rays are produced by 10 TeV protons, then since $f(> 10 TeV) \sim 1$ we require $\epsilon_p \epsilon_A > 0.1$ to reproduce the long term low flux at this energy. Even with $\epsilon_p \epsilon_A = 1$, it is not possible to account for the burst flux. At the higher energy, $\gamma$-rays of 100 TeV require parent protons near 1000 TeV, where $f(> 1000 TeV)$ is only $10^{-3}$, and even the low fluxes reported are only explained if $\epsilon_p \epsilon_A = 1$.

Clearly, the observed luminosity of the $\gamma$-rays is still a problem. Either the beaming factor of the protons needs to be larger than found in this simulation, or the power available from accretion must occasionally be much greater than the average bolometric luminosity. Since the latter is already Eddingtonian, we require occasional super-Eddington accretion rates to explain the observations. Such a situation is not so unrealistic since the accretion geometry is decidely asymmetric. (This requirement is true of course for any model which ultimately draws its power from the gravitational infall of matter, and the extent to which we must appeal to super-Eddingtonian accretion is much less in a model like this where the protons are naturally beamed.) An implication of increased accretion rates is that the gas in the accretion column will be accordingly denser, which will improve the production of $\gamma$-rays in the column itself up to a point, but ultimately $pp$ collisions will degrade the maximum energy achievable.

## 8. Conclusions

A simple model for the acceleration of protons by scattering off relativistic Alfvén waves has been set up in the well understood environment of the accretion column of Her X-1. It may be applicable to other similar accreting X-ray pulsars and possibly some cataclysmic variables for which TeV emission has also been detected, but Her X-1 remains the prototype of this type of source and the only one for whose parameters are very well established. The mechanism is reminiscent of classical second-order Fermi acceleration, except that here the "magnetic clouds" are moving relativistically. When plausible parameters are assumed for the spectrum of Alfvén waves and the initial state of the particles, most protons are accelerated to energies in excess of 10 TeV, with a mean energy of 30 TeV, and an $\varepsilon^{-2}$ spectrum which extends beyond 100 TeV before steepening. Particle injection slightly nearer to the neutron star shifts the mean proton energy to around 200 TeV, and uniform injection of protons along the column would raise the mean energy to this value. The maximum energy seems to be mainly limited by the maximum wavelength of the Alfvén waves. The presence of waves with wavelength $\sim R_n$ enhances the mean energy to 250 TeV.

The highest individual energy achieved is over 1 PeV, limited only by the statistical nature of the simulation, although the spectrum has steepened markedly by this point. Whilst mean energies of over 100 TeV are achieved in favourable circumstances, 30 TeV seems achievable following reasonable assumptions about the environment, and production of 1 TeV protons seems to be an almost inevitable by-product of the accretion mechanism in this system.

For the simulation, the single test-particle approach has been followed. No attempt has been made, however, to consider the energetics of a flow of such protons: how they interact back on the system, and exactly what luminosity can be supported in such a flow are very complex questions that require further work. Nor has the question of particle injection been addressed. In the simulation, relativistic protons were injected into the mid-point of the accretion column, but of course the injection could occur at any point or even uniformly along the column. The mechanism for particle injection is yet another question. Here we simply assume that some of the relativistic particles produced in the accretion process find their way into the upper reaches of the column.

The process has the advantage of producing a focussed beam of protons which substantially reduces the power requirement of high energy particles. The beam emerges in the direction of the inner accretion disk, where there is sufficient material for the production of $\gamma$-rays, and which may help explain the blue-shifted nature of the pulsed emission [45]. Explaining the flux of observed $\gamma$-rays is still a problem however, even with the beaming. A pulsed signal of other secondaries is possible, including neutrinos. Sporadic detection of secondaries is easily explained if the beam wanders around for some reason e.g. due to the precession of the nuetron star and hence the magnetic field. Constant low-level production of such secondaries is also possible, due to the thin gas embedded in the accretion column itself.

*Acknowledgements*
This work was undertaken with the aid of an SERC grant. I gratefully thank A. M. Hillas for many useful discussions whilst this work was underway, and R. J. Protheroe and an anonymous referee for useful comments on the manuscript.